\documentclass{IEEE_lsens}
\usepackage[noadjust]{cite}
\ifCLASSINFOpdf
\else
\fi
\usepackage[T1]{fontenc}
\usepackage[cmintegrals]{newtxmath}
\usepackage{bm}
\usepackage{array}
\usepackage{url}
\usepackage{cite}
\usepackage{amsmath,amsfonts}
\usepackage{graphicx}
\usepackage{textcomp}
\usepackage{xcolor}
\usepackage[ruled]{algorithm2e}
\usepackage{balance}
\usepackage{wrapfig}
\usepackage{float}
\usepackage{subfig}
\usepackage{times}
\usepackage{doi}
\usepackage{hyperref}
\usepackage{caption}
\usepackage{multirow}

\ifCLASSINFOpdf
\else
\fi
\providecommand{\hypersetup}[1]{\relax}

\hypersetup{pdftitle={CWT-Enhanced Vibration Sensing With Time-Frequency Region Localization Using YOLO},
pdfsubject={Typesetting},
pdfauthor={Po-Heng Chou},
pdfkeywords={Class, IEEE, IEEE\_lsens, IEEE Sensors Letters, LaTeX, Typesetting, TeX}}

\begin{document}
\markboth{Vol.~X, No.~X, XXX~2026}{0000000}
\IEEELSENSarticlesubject{Sensor Applications}%
\title{CWT-Enhanced Vibration Sensing With Time-Frequency Region Localization Using YOLO}
\author{\IEEEauthorblockN{Po-Heng Chou\IEEEauthorrefmark{1}\IEEEauthorieeemembermark{1}, Wei‑Lung Mao\IEEEauthorrefmark{2}, Ru-Ping Lin\IEEEauthorrefmark{2}, Jen-Yu Chiu\IEEEauthorrefmark{2}, and Chun-Yu Yeh\IEEEauthorrefmark{2}}
\IEEEauthorblockA{\IEEEauthorrefmark{1}Research Center for Information Technology Innovation (CITI), Academia Sinica (AS), Taipei 11529, Taiwan\\
\IEEEauthorrefmark{2}Department of Electrical Engineering, National Yunlin University of Science and Technology (NYUST), Yunlin 64002, Taiwan\\
\IEEEauthorieeemembermark{1}Member, IEEE\\
}
\thanks{Corresponding author: Po-Heng Chou (e-mail: d00942015@ntu.edu.tw).\protect}
\thanks{Associate Editor: .}%
\thanks{Digital Object Identifier 10.1109/LSENS.2026.0000000}}

\IEEELSENSmanuscriptreceived{\vspace{-0.2in}Manuscript received May XX, 2026;
revised XXX XXX, 2026; accepted XXX XXX, 2026.
Date of publication XXX XXX, 2026; date of current version XXX XXX, 2026.}

\IEEEtitleabstractindextext{%

\begin{abstract}
\begin{wrapfigure}{r}{0.45\textwidth}
\hspace{-10pt}
\includegraphics[width=0.45\textwidth]{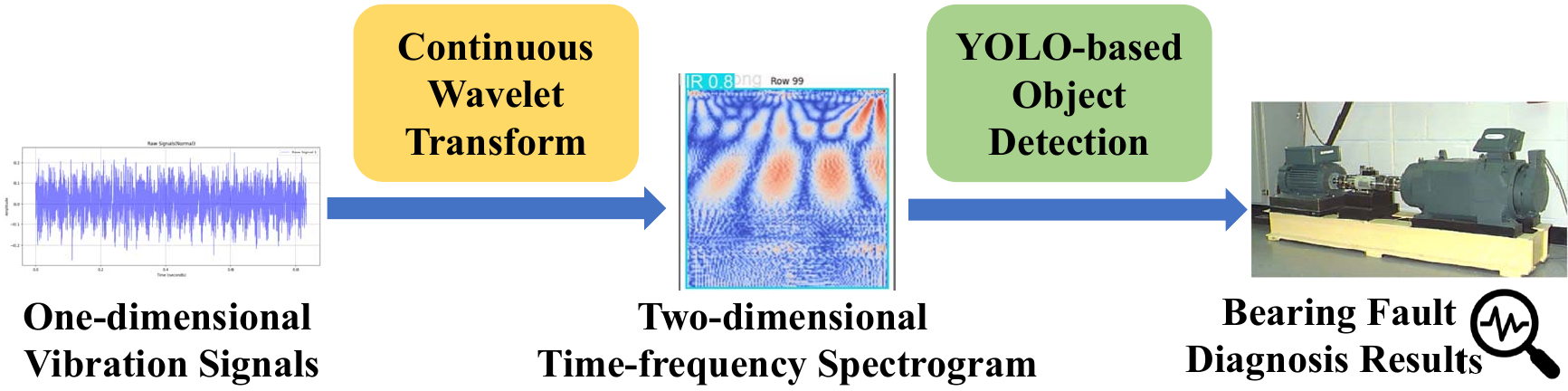}
\vspace{-5pt}
\end{wrapfigure}
This letter presents a CWT-enhanced vibration sensing framework for bearing fault monitoring through localized time-frequency region detection on continuous wavelet transform (CWT) spectrograms. Vibration signals are transformed into CWT spectrograms to improve the observability of weak and non-stationary fault signatures, and YOLOv9, YOLOv10, and YOLOv11 are employed to detect and identify localized fault-related energy regions in the time-frequency domain. Experiments on the CWRU, PU, and IMS datasets show that the proposed framework improves the detectability and robustness of fault-related sensing patterns compared with conventional time-series models, modern vision backbones, and short-time Fourier transform (STFT)-based representations, achieving mean average precision (mAP) values up to 99.4\%, 97.8\%, and 99.5\%, respectively. In addition, the localized region detection framework provides a more interpretable relationship between time-frequency energy distributions and characteristic bearing fault frequencies. {\color{black}These results demonstrate an effective and generalizable approach for interpretable vibration sensing in noisy industrial environments.}
\end{abstract}

}

\maketitle

\section{Introduction}

Rolling bearings are critical components in rotating machinery, and operational reliability directly influences equipment lifespan, production efficiency, and safety~\cite{Zhang2020_DLReview}. Studies suggest that nearly 40\% of rotating machinery failures originate from bearing faults~\cite{Neupane2020_Review}. In practical rotating systems, bearing health is monitored through vibration sensing, where transducers capture one-dimensional time-domain signals reflecting the mechanical condition of the machine. However, early-stage fault signatures are often weak, transient, and highly non-stationary, making them difficult to observe and interpret under realistic operating conditions~\cite{Zhang2020_DLReview,Neupane2020_Review}.

Vibration signals acquired from rotating machinery contain impulsive features related to mechanical defects, such as inner-race, outer-race, or ball faults~\cite{Xia2018_CNN_Multisensor}. Nevertheless, raw time-domain waveforms do not reveal how fault-related energy is distributed across time and frequency, making it difficult to distinguish fault types with similar temporal behavior but different spectral characteristics~\cite{Neupane2020_Review}.

Several classical time-frequency analysis methods, such as short-time Fourier transform (STFT), Wigner-Ville distribution (WVD)~\cite{WVD}, and Hilbert-Huang transform (HHT)~\cite{EMD_HHT}, have been widely adopted to characterize non-stationary vibration signals. However, these approaches suffer from inherent limitations, such as fixed time-frequency resolution, cross-term interference, and sensitivity to noise, which restrict their effectiveness in complex sensing environments~\cite{Zhang2020_DLReview,Neupane2020_Review}. In contrast, continuous wavelet transform (CWT)~\cite{Morelt}, particularly with Morlet wavelets, provides multi-resolution analysis with better localization in both time and frequency. Its ability to highlight transient and scale-varying patterns makes it especially effective for exposing early fault signatures that are otherwise difficult to observe in raw vibration measurements~\cite{Patil2024}.

Beyond signal representation, the subsequent interpretation of sensing patterns is critical. In recent years, deep learning (DL) methods have demonstrated strong performance in bearing fault diagnosis due to their end-to-end feature learning capability~\cite{Xia2018_CNN_Multisensor,Pan2018_CNNLSTM,Zhang2020_DLReview,Neupane2020_Review}. 
Despite recent advances in DL-based fault diagnosis, most existing methods rely on global classification of entire signal segments, which may dilute localized fault-related transient structures in the time-frequency domain. This motivates localized time-frequency region detection for vibration sensing analysis.
Based on this insight, this work reformulates vibration-based bearing fault monitoring as a localized time-frequency region detection problem on time-frequency representations. Specifically, fault-induced transients in CWT spectrograms appear as localized energy clusters associated with characteristic fault-related frequency components. This property enables the identification of localized fault-related sensing features through region-based detection. Accordingly, we adopt You Only Look Once (YOLO)-based detectors, including YOLOv9~\cite{wang2025yolov9}, YOLOv10~\cite{wang2024yolov10}, and YOLOv11~\cite{ultralytics2025yolov11}, to identify these localized regions in CWT spectrograms. Unlike conventional classifiers, YOLO provides region-aware predictions that preserve the structural distribution of fault-related energy patterns in the time-frequency domain, thereby improving both interpretability and diagnostic precision.
It should be noted that the proposed framework localizes fault-related energy concentrations in the time-frequency spectrogram domain rather than performing direct physical-space localization of bearing defects.

The main contributions of this letter are as follows:

\begin{itemize}
    \item We reformulate vibration-based bearing fault monitoring as a localized time-frequency energy-region detection problem, enabling the identification of fault-related transient patterns from vibration measurements.

    \item We propose a CWT-enhanced sensing framework that improves observability of weak and non-stationary fault signatures through multi-resolution time-frequency analysis and localized time-frequency region detection.

    \item We establish an interpretable connection between
    localized regions in CWT spectrograms and characteristic fault-related frequency components, providing insight into fault-related transient energy distributions beyond conventional black-box classification.

    \item Extensive experiments on the CWRU~\cite{CWRU}, PU~\cite{PU}, and IMS~\cite{IMS} datasets demonstrate that the proposed framework improves the effective sensing capability, including enhanced detectability and robustness of fault-related signatures, compared with representative time-series models~\cite{Chen2021_MCNNLSTM}, modern vision backbones~\cite{liu2022convnet,Song2024_ConvNeXt}, and alternative time-frequency representations such as STFT.
\end{itemize}

\section{System Overview and Proposed CWT-YOLO}

The proposed framework is designed from a vibration-sensing perspective, where the objective is to enhance the observability and interpretability of fault-related signatures embedded in sensor measurements. The overall architecture consists of three main stages: (1) time-frequency transformation using CWT, (2) spectrogram pre-processing and labeling, and (3) localized time-frequency region detection using YOLO-based object detectors.
Representative vibration signals are transformed into
time-frequency spectrograms using CWT for subsequent
fault-region detection.

\vspace{-0.1in}
\subsection{Continuous Wavelet Transform (CWT)}

The CWT is a time-frequency analysis tool that decomposes a one-dimensional signal into localized time-scale components using a family of wavelets. Compared to fixed-window methods like the STFT, CWT enables multi-resolution analysis with dynamic windowing, making it especially suitable for analyzing non-stationary signals such as bearing vibrations. The CWT of a signal \( x(t) \) is defined as:
\begin{equation}
    \text{CWT}(a, b) = \frac{1}{\sqrt{|a|}} \int_{-\infty}^{\infty} x(t) \, \psi^*\left( \frac{t - b}{a} \right) dt,
    \label{eq:cwt}
\end{equation}
where \( a \) denotes the scale (inverse frequency), \( b \) is the time shift, \( \psi(t) \) is the mother wavelet, and \( \psi^*(t) \) its complex conjugate.

In this work, the Morlet wavelet is adopted due to its favorable joint time-frequency localization capability, balanced multi-resolution analysis across different frequency bands, and high sensitivity to transient impulsive components commonly observed in bearing fault signals~\cite{Patil2024}. The Morlet wavelet combines a sinusoidal carrier with a Gaussian envelope, $\psi(t)=\pi^{-1/4}e^{j\omega_0 t}e^{-t^2/2}$, where $\omega_0$ denotes the central frequency. In addition, the scale range was selected according to the characteristic frequency distributions of the bearing datasets to ensure adequate coverage of fault-related transient components.
By applying CWT to raw vibration signals, we generate two-dimensional time-frequency spectrograms that expose transient features and scale-dependent energy distributions. The generated spectrograms expose localized transient energy distributions suitable for subsequent region-based fault detection.

\vspace{-0.1in}
\subsection{Time-Frequency Spectrogram Visualization}

\begin{figure}[t]
\centering
\subfloat[Inner race]{
\includegraphics[width=0.45\linewidth]{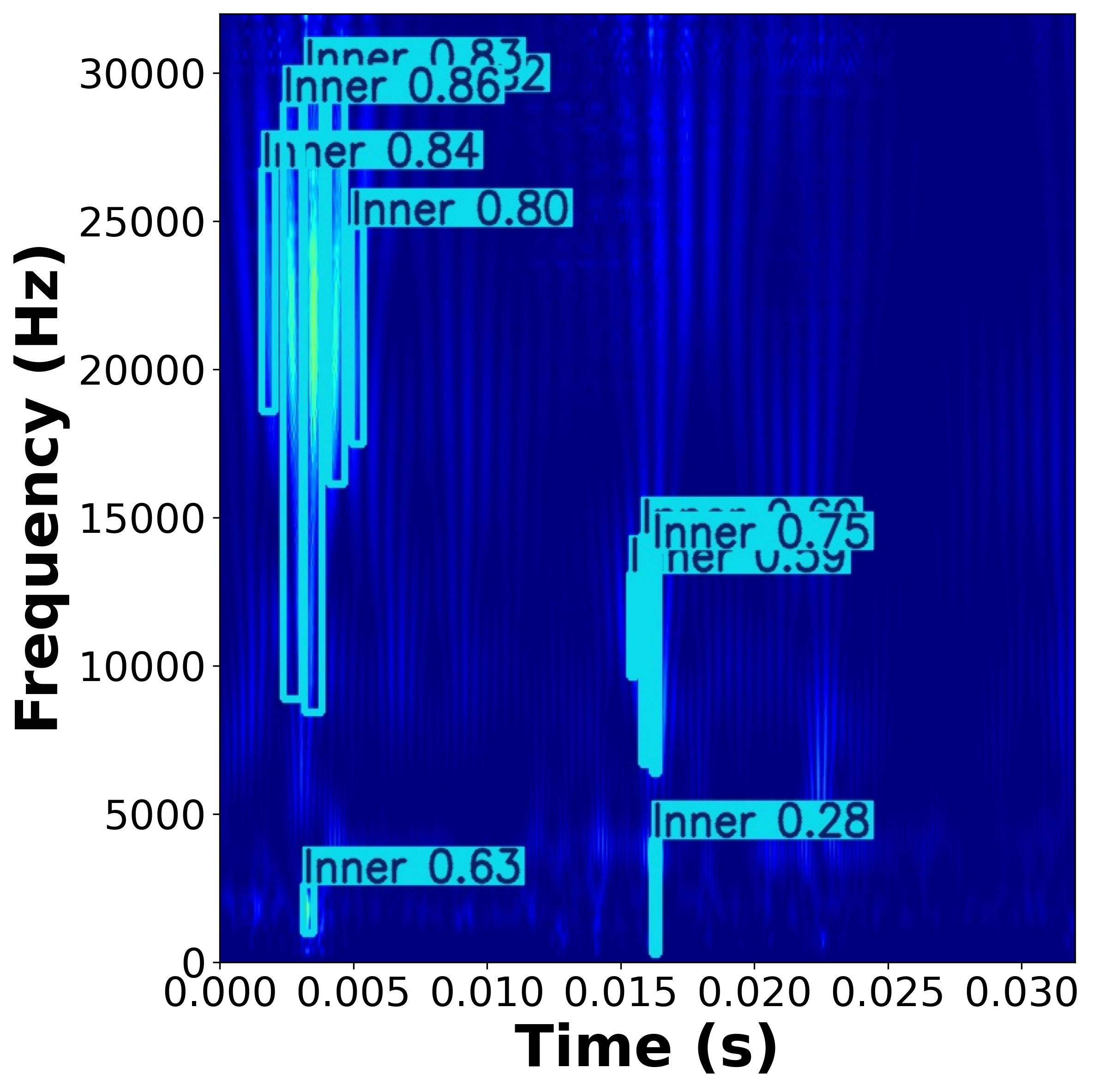}}
\hfill
\subfloat[Outer race]{
\includegraphics[width=0.45\linewidth]{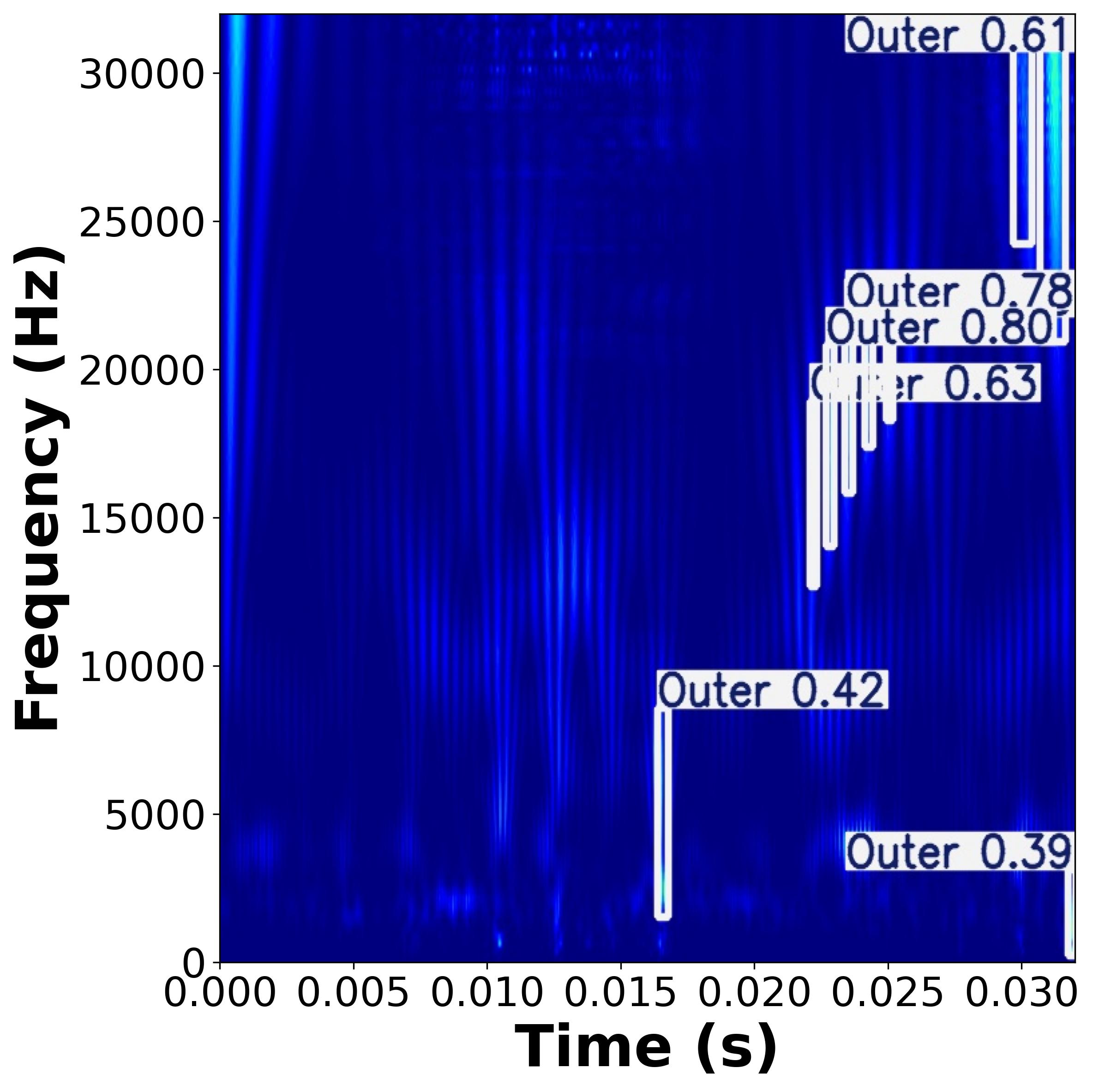}}

\caption{Representative CWT spectrograms illustrating localized fault-related energy distributions under different bearing fault conditions.}
\label{fig:spectrogram}
\vspace{-0.1in}
\end{figure}

To validate the visual enhancement provided by CWT, we convert the same vibration signals into two-dimensional time-frequency spectrograms using Morlet-based wavelet transformation. The resulting images, shown in Fig.~\ref{fig:spectrogram}, capture distinct energy concentration patterns that correspond to different fault types.

\vspace{-0.05in}
\subsection{Data Pre-processing and Labeling}

To ensure that the CWT-generated spectrograms are suitable for training YOLO-based object detectors, a series of pre-processing and annotation steps is applied to the vibration data prior to model inference.

\subsubsection{Signal Segmentation and Spectrogram Generation}

To prevent data leakage, the dataset is first partitioned at the raw signal level before segmentation, ensuring that segments derived from the same original signal do not appear across training, validation, and test sets, even under overlapping window settings.

Each segment is transformed into a 2D time-frequency spectrogram using the Morlet-based CWT described earlier. The resulting grayscale spectrograms are logarithmically compressed and normalized to a $[0, 1]$ range.
All spectrograms are resized to $640 \times 640$ pixels to meet the input specification of YOLOv9, v10, and v11 models.

\subsubsection{Bounding Box Annotation for Object Detection}

Unlike traditional classification models that produce a single label for an entire signal segment, YOLO models require localized object annotations in the form of bounding boxes. {\color{black}The bounding box annotation followed a threshold-assisted manual procedure. Each normalized CWT spectrogram was first binarized using a fixed normalized intensity threshold
determined separately for each dataset to generate candidate high-energy connected components. These candidates were then refined in LabelImg by two annotators according to the same guideline: enclosing compact impulsive energy clusters distinguishable from the background and consistent with fault-related transient patterns. Disagreements were resolved by consensus inspection, and the final boxes were used as ground-truth labels for YOLO training and evaluation.} Each box is associated with a class label corresponding to one of four fault types: Normal, Ball fault, Inner race fault, and Outer race fault.
The labeled regions correspond to interpretable
fault-related energy concentrations in the
time-frequency domain.
These annotations are saved in YOLO format, specifying each object as a row of normalized values \texttt{[class\_id, x\_center, y\_center, width, height]} with respect to the image dimensions. Future work may explore semi-automated labeling using energy-based region proposal algorithms to further reduce human bias.

\subsubsection{Dataset Splitting and Augmentation}

Labeled data are partitioned into training, validation, and test sets using an $80:10:10$ ratio to ensure sufficient coverage and generalization. To enhance model robustness, data augmentation techniques including horizontal flipping, small-scale rotation $\pm5^\circ$, and contrast jittering are applied.

\vspace{-0.11in}
\subsection{YOLO-based Fault Detection}

In this study, vibration-based fault sensing is reformulated as a localized time-frequency region detection problem on CWT-generated spectrograms.
This reformulation enables the sensor measurements to be interpreted in terms of localized energy distributions, thereby improving the observability of weak and transient fault signatures.
Rather than assigning a global label to each signal segment, the system identifies and localizes regions of interest (ROIs) where fault-related energy patterns occur, enabling both classification and localized time-frequency interpretation.

{\color{black} As described in Section II-C, binarization generated candidate ROIs, while the final ground-truth boxes were manually verified and refined by consensus.} By directly mapping pixel intensity to signal energy magnitude, this approach effectively isolates peak energy clusters while filtering out background noise~\cite{SUZUKI198532}.

Crucially, these ROIs detected by YOLO correlate with localized fault-related energy concentrations associated with characteristic bearing fault frequencies. For example, the detected regions correlate with characteristic fault-related frequency components, such as BPFI and its harmonics.

We adopt YOLOv9~\cite{wang2025yolov9},
YOLOv10~\cite{wang2024yolov10}, and
YOLOv11~\cite{ultralytics2025yolov11}
due to their strong detection capability and favorable
efficiency-accuracy trade-off for localized
time-frequency region detection. 
All models are trained on $640 \times 640$ spectrograms derived from the CWRU~\cite{CWRU}, PU~\cite{PU}, and IMS~\cite{IMS} datasets, using bounding box annotations over energy-dense fault regions. A composite loss function including localization, objectness, and classification terms guides the training. In addition, all YOLO variants are trained under the same input resolution, augmentation strategy, and optimization settings to ensure fair comparison across models.

\begin{figure}[t] 
\centering 
\includegraphics[width=0.7\linewidth]{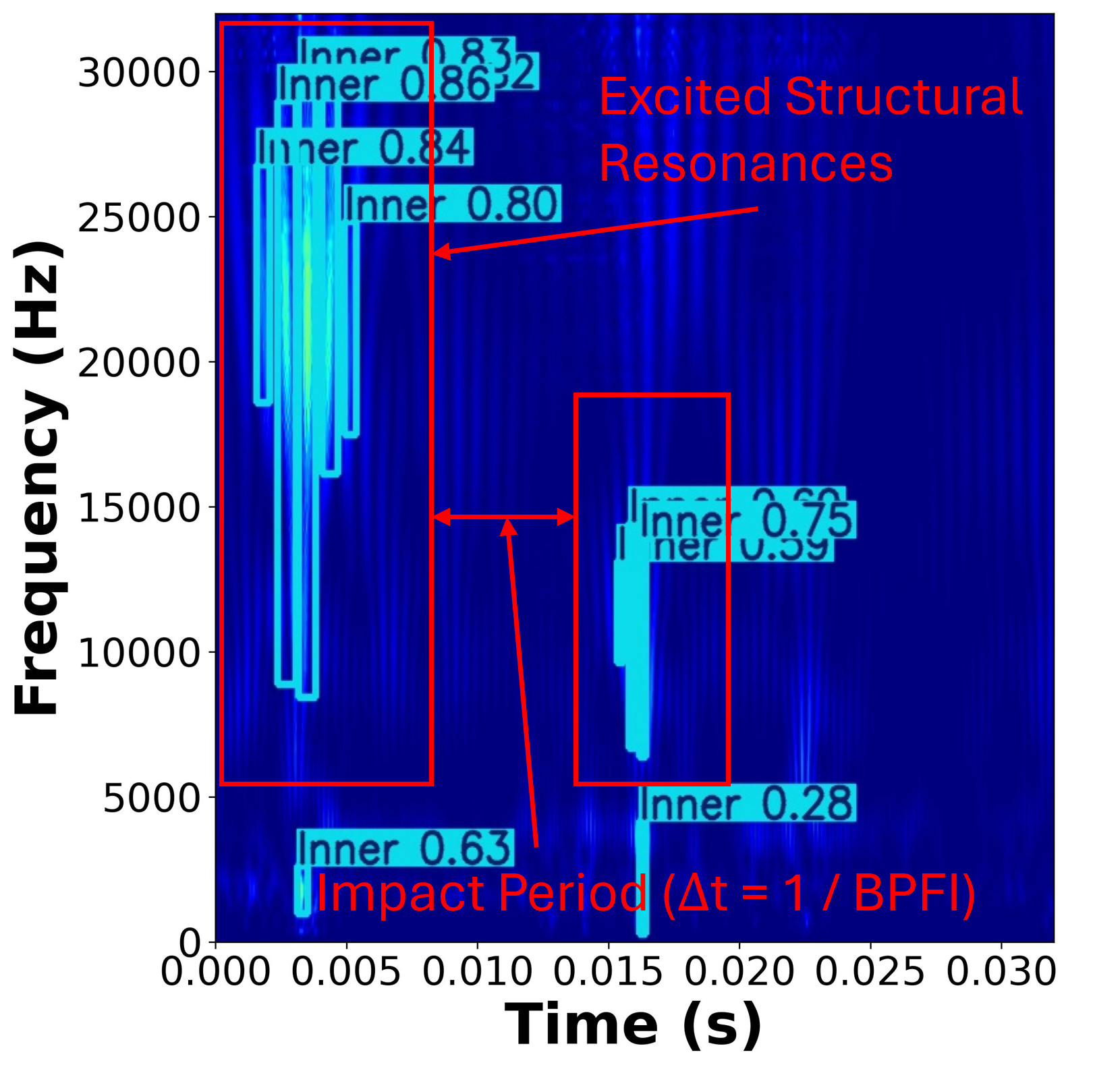} 
\caption{\textcolor{black}{
Time-frequency interpretation of YOLO-detected regions on CWT spectrograms (inner race fault). The boxes capture high-energy transient bursts, while their temporal spacing ($\Delta t$) matches the characteristic fault period (e.g., $1/\text{BPFI}$; BPFI: ball pass frequency of the inner race). This shows that the framework captures interpretable fault-related energy distributions rather than generic global patterns.
}}
\label{fig:localization_evidence}
\vspace{-0.1in}
\end{figure}

\vspace{-0.2in}
\section{Experimental Results}



\subsection{Experimental Setup}

All models are trained and evaluated on three public bearing fault datasets: CWRU~\cite{CWRU}, PU~\cite{PU}, and IMS~\cite{IMS}. The CWT spectrograms are generated from raw vibration signals with 2048-sample windows and $50\%$ overlap. Images are resized to $640 \times 640$ pixels and annotated with bounding boxes using the LabelImg tool. The datasets are partitioned into 80\% training, $10\%$ validation, and $10\%$ testing splits. All DL models are trained for $500$ epochs using stochastic gradient descent (SGD) (learning rate=0.01, batch size=8).

\vspace{-0.1in}
\subsection{Performance Comparison Across Datasets}

Table~\ref{tab:performance} presents the performance comparison across different feature representations and models. 
YOLO-based models with CWT consistently achieve superior performance across all datasets, demonstrating improved detectability of fault-related sensing patterns and outperforming both traditional 1D-based methods (MCNN-LSTM~\cite{Chen2021_MCNNLSTM}) and modern vision backbones (ConvNeXt~\cite{Song2024_ConvNeXt}).
In contrast, replacing CWT with STFT leads to substantial performance degradation for all YOLO variants, particularly on the PU dataset, highlighting the importance of multi-resolution feature representation for capturing localized transient responses.
Among all configurations, YOLOv11 with CWT achieves the best overall performance, demonstrating strong robustness under varying operating conditions and improved sensitivity to weak fault signatures.

To evaluate computational efficiency, Table~\ref{tab:flops_params} summarizes the theoretical FLOPs and parameter counts of the compared models. Compared to YOLOv9, YOLOv11 reduces FLOPs by over 97\% and parameters by nearly 95\%, while achieving comparable or better accuracy.
Although MCNN-LSTM achieves the lowest computational cost, its performance is significantly inferior to YOLO-based models in terms of detection accuracy and robustness.
YOLOv11 provides an optimal trade-off between efficiency and high detection accuracy, making it highly suitable for lightweight industrial fault monitoring and practical sensing deployment scenarios.

\begin{table}[t]
\centering
\footnotesize
\caption{Performance comparison across different feature representations and models on CWRU, PU, and IMS datasets.}
\label{tab:performance}
\resizebox{1\linewidth}{!}{
\begin{tabular}{|c|c|c|c|c|c|c|}
\hline
\textbf{Dataset} & \textbf{Feature} & \textbf{Model} & \textbf{mAP@0.5} & \textbf{PRE} & \textbf{REC} & \textbf{F1} \\
\hline

\multirow{8}{*}{CWRU~\cite{CWRU}} 

& \multirow{4}{*}{\textbf{CWT}} 
& YOLOv9 & \textbf{99.4\%} & 98.6\% & \textbf{98.5\%} & \textbf{98.6\%} \\
& & YOLOv10 & \textbf{99.4\%} & \textbf{99.2\%} & 98.1\% & \textbf{98.6\%} \\
& & YOLOv11 & 99.0\% & 93.9\% & \textbf{98.5\%} & 96.2\% \\
& & ConvNeXt~\cite{Song2024_ConvNeXt} & 97.0\% & 98.2\% & 96.7\% & 97.3\% \\

\cline{2-7}

& \multirow{3}{*}{STFT} 
& YOLOv9 & 86.5\% & 81.8\% & 77.4\% & 79.5\% \\
& & YOLOv10 & 77.2\% & 73.5\% & 69.6\% & 71.5\% \\
& & YOLOv11 & 78.5\% & 73.5\% & 71.8\% & 72.6\% \\

\cline{2-7}

& Raw (1D) 
& MCNN-LSTM~\cite{Chen2021_MCNNLSTM} & 96.0\% & 96.1\% & 96.1\% & 96.1\% \\

\hline

\multirow{8}{*}{PU~\cite{PU}} 

& \multirow{4}{*}{\textbf{CWT}} 
& YOLOv9 & 91.6\% & 80.8\% & 84.8\% & 82.7\% \\
& & YOLOv10 & 97.2\% & 89.0\% & 92.7\% & 90.8\% \\
& & YOLOv11 & \textbf{97.8\%} & \textbf{94.9\%} & \textbf{93.8\%} & \textbf{94.3\%} \\
& & ConvNeXt~\cite{Song2024_ConvNeXt} & 50.4\% & 59.9\% & 50.4\% & 50.7\% \\

\cline{2-7}

& \multirow{3}{*}{STFT} 
& YOLOv9 & 26.6\% & 23.3\% & 36.6\% & 28.5\% \\
& & YOLOv10 & 35.5\% & 34.8\% & 40.6\% & 37.5\% \\
& & YOLOv11 & 32.9\% & 29.3\% & 41.6\% & 34.4\% \\

\cline{2-7}

& Raw (1D) 
& MCNN-LSTM~\cite{Chen2021_MCNNLSTM} & 77.7\% & 77.7\% & 77.4\% & 77.6\% \\

\hline

\multirow{8}{*}{IMS~\cite{IMS}} 

& \multirow{4}{*}{\textbf{CWT}} 
& YOLOv9 & \textbf{99.5\%} & 99.9\% & \textbf{100.0\%} & \textbf{100.0\%} \\
& & YOLOv10 & \textbf{99.5\%} & 99.9\% & \textbf{100.0\%} & 99.9\% \\
& & YOLOv11 & \textbf{99.5\%} & \textbf{100.0\%} & \textbf{100.0\%} & \textbf{100.0\%} \\
& & ConvNeXt~\cite{Song2024_ConvNeXt} & 100.0\% & 100.0\% & 100.0\% & 100.0\% \\

\cline{2-7}

& \multirow{3}{*}{STFT} 
& YOLOv9 & 45.9\% & 53.8\% & 43.4\% & 48.0\% \\
& & YOLOv10 & 41.6\% & 46.9\% & 38.7\% & 42.4\% \\
& & YOLOv11 & 48.9\% & 50.9\% & 46.7\% & 48.7\% \\

\cline{2-7}

& Raw (1D) 
& MCNN-LSTM~\cite{Chen2021_MCNNLSTM} & 96.8\% & 96.8\% & 96.8\% & 96.8\% \\

\hline
\end{tabular}
}
\end{table}

\begin{table}[t]
\centering
\scriptsize
\caption{Comparison of model complexity and inference performance (measured on NVIDIA RTX 3080 Ti).}
\label{tab:flops_params}
\resizebox{1\linewidth}{!}{
\begin{tabular}{|c|c|c|c|c|}
\hline
\textbf{Model} & \textbf{FLOPs (G)} & \textbf{Params (M)} & \textbf{Inference (ms)} & \textbf{FPS} \\
\hline
YOLOv9      & 236.7 & 48.35 & 15.24 & 57.40 \\
YOLOv10     & 8.2   & 2.57  & 11.54 & 77.42 \\
YOLOv11     & 6.3   & 2.46  & 11.43 & 73.72 \\
MCNN-LSTM~\cite{Chen2021_MCNNLSTM}  & \textbf{0.010}  & \textbf{0.352} & \textbf{2.15} & \textbf{465.12} \\
ConvNeXt~\cite{Song2024_ConvNeXt} & 4.5 & 28.6 & 7.80 & 128.21 \\
\hline
\end{tabular}
}
\vspace{-0.1in}
\end{table}

\subsection{Ablation Study on Time-Frequency Features}

\begin{figure}[t] 
    \centering 
    \subfloat[]{\includegraphics[width=0.28\linewidth]{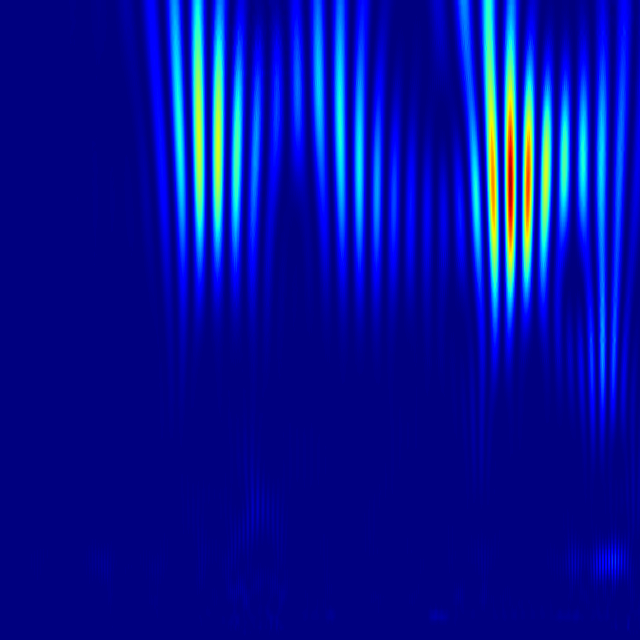}} 
    \hspace{+1mm}
    \subfloat[]{\includegraphics[width=0.28\linewidth]{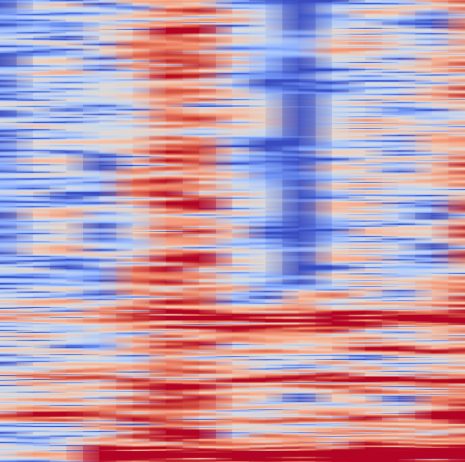}} 
    \caption{Visual comparison of time-frequency features for an inner race fault from the PU dataset: (a) CWT and (b) STFT spectrograms.} 
    \label{fig:feature_comparison}
    \vspace{-0.2in}
\end{figure}

\vspace{-0.1in}
To validate the necessity of CWT, an ablation study comparing it with STFT was conducted. For fair comparison, the STFT spectrograms were generated using a Hamming window with a 256-point window length, 50\% overlap, and 512-point FFT resolution. As shown in Table~\ref{tab:performance}, switching to STFT causes significant performance degradation across all datasets, most notably on PU (97.8\% to 32.9\%). This degradation is attributed to the fixed time-frequency resolution of STFT, which blurs transient energy distributions and reduces the observability of localized fault signatures. In contrast, CWT enables sharp localization of fault signatures across multiple scales, thereby improving the effective sensing capability in detecting weak and non-stationary fault patterns.

\vspace{-0.01in}
To further evaluate the robustness of the proposed method under harsh industrial environments, Additive White Gaussian Noise (AWGN) was introduced to the raw vibration signals of the IMS test set at varying signal-to-noise ratio (SNR) levels. As shown in Table~\ref{tab:snr}, the proposed CWT-YOLOv11 framework consistently outperforms the STFT-based representation under low-SNR conditions. This significant improvement is mainly attributed to the multi-resolution property of CWT, which preserves localized transient energy distributions more effectively under heavy background interference, whereas STFT features become severely obscured.

\begin{table}[H]
\centering
\caption{\textcolor{black}{Noise robustness comparison under different SNR conditions on the IMS test set.}}
\label{tab:snr}
\resizebox{0.8\linewidth}{!}{
\begin{tabular}{|c|c|c|}
\hline
\textbf{SNR Level} & \textbf{CWT-YOLOv11 (mAP)} & \textbf{STFT-YOLOv11 (mAP)} \\
\hline
Original (No Noise) & 99.5\% & 48.9\% \\
10 dB & 68.0\% & 2.3\% \\
5 dB  & 52.5\% & 0.1\% \\
0 dB  & 27.9\% & Failed (≈ 0\%) \\
\hline
\end{tabular}
}
\vspace{-0.1in}
\end{table}

As observed in Table~\ref{tab:snr}, the CWT-YOLOv11 model demonstrates strong noise robustness under harsh
conditions. While the mAP naturally degrades as the noise power increases, the model successfully maintains
68.0\% and 52.5\% mAP under 10dB and 5dB SNR,
respectively. Furthermore, the inner race fault features
remain partially detectable even under 0dB extreme noise,
demonstrating that CWT preserves transient fault-related
energy distributions more effectively than STFT under
heavy background interference.
{\color{black}At 0 dB, false detections, missed impulses, or boxes violating expected fault-periodic spacing may occur because YOLO does not embed mechanical periodicity constraints.
}

\section{Conclusion}
This letter presents a CWT-enhanced vibration sensing framework with a YOLO-based time-frequency region
localization for spectrogram analysis. Evaluations on the CWRU, PU, and IMS datasets confirm improved detectability and robustness of fault-related signatures over conventional methods. These results show that localized time-frequency region detection on CWT spectrograms provides an effective and interpretable approach {\color{black}for vibration-based fault monitoring in noisy and complex industrial environments.}

\vspace{-0.1in}
\bibliography{IEEE_Double}
\bibliographystyle{IEEEtran}
\end{document}